**Structure of Working Memory in Children From 3 to 8 Years Old**

Barbara Carretti*, David Giofrè§, Enrico Toffalini*, Cesare Cornoldi*, Massimiliano Pastore° e Silvia Lanfranchi°

*Department of General Psychology, University of Padova, Italy

§Department of Educational Sciences, University of Genova, Italy

°Department of Developmental Psychology and Socialization, University of Padova, Italy

**Author note**

# Structure of Working Memory in Children From 3 to 8 Years Old

## Abstract

Several models of working memory (WM) have been proposed in the literature. Most of the research on the architecture of WM is based on adults or older children, but less is known about younger children. In this study, we tested various models of WM on a sample of 739 Italian children from 3 to 8 years old. Participants were assessed with 12 WM tasks, systematically varying the modality and level of executive control required (based on the number of activities to be performed at once: retention alone, ignoring distractors, and dealing with dual tasks). We examined younger children, $n = 501$, $M_{age} = 56.8$ months ($SD = 6.4$, 48% males) and older children, $n = 238$, $M_{age} = 80.0$ months ($SD = 9.0$, 58% males) separately using multigroup confirmatory factor analyses. A Bayesian analytical approach was adopted. Our results suggested that a four-factor model distinguishing between verbal, visual, spatial-simultaneous, and spatial-sequential components of WM achieved the best fit. Overall, the WM structure was very similar in the two groups. We further explored this result with an additional model with a central executive factor loaded on high-control tasks only, and found evidence for the presence of an executive control component. The contribution of this factor in terms of explained variance was only modest, however. Our findings demonstrate that it is important to distinguish between WM components in young children.

**Structure of Working Memory in Children From 3 to 8 Years Old**

Working memory (WM) is a limited-capacity system that allows for the storage and manipulation of information over a brief period of time. Several models of WM have been described in the literature (see Cornoldi & Giofré, 2014, for a review). Some models distinguish between the ways they present the information to be processed; these are called domain-specific or modality-specific models. Others stress the importance of attentional control over the content of the information to be processed and are known as domain-general or modality-independent models.

Domain-specific models generally refer to the distinction between two slave systems (i.e., the speech-based phonological loop and the visuospatial sketchpad), which are included in the classical model proposed by Baddeley and Hitch (1974). Their so-called tripartite model includes both domain-specific and domain-general components; the domain-specific components are controlled by the central executive (CE), an attentional control system of limited capacity (Baddeley, 2000; Baddeley & Hitch, 1974; Shah & Miyake, 1996). Speculating on Baddeley's (2000) proposal, Shah and Miyake (1996) proposed an alternative model, suggesting that WM is supported by two separate pools of domain-specific resources for verbal and visuospatial information. Each domain is independently capable of manipulating information and keeping it active, although the information includes domain-general resources as well.

From a different perspective, domain-general models stress the importance of attentional control as the key aspect of WM (e.g., Cowan, 2001; Engle et al., 1999). In this case, WM is seen as a mental workspace in which only one consistent content can be dominant at any given moment (Vergauwe & Cowan, 2014). Such models distinguish only between tasks requiring less cognitive control - often called short-term memory (STM) tasks - and those demanding more cognitive control, called WM or complex-span tasks (see Engle, 2010, for an overview).

Other models that combine the mentioned approaches have been proposed as well. For example, the Cornoldi and Vecchi's (2003) model envisages WM (and WM tasks) as lying along two continua that describe the type of content to be processed (horizontal continuum) and the degree of active control demanded by the task (vertical continuum). As for the horizontal continuum, the model distinguishes not only between verbal and visuospatial components, but also within the visuospatial sketchpad. This is in line with other accounts suggesting different mechanisms underlying visual and spatial memory (e.g., Logie, 1995; Pickering, 2001) and existing between spatial-simultaneous and spatial-sequential processes (e.g., Mammarella et al., 2006). The latter distinction refers to the format of presentation of the stimuli, with the items to be recalled presented simultaneously in spatial-simultaneous tasks (termed "static" by Gathercole & Pickering, 2001) or one at a time in spatial-sequential tasks (termed "dynamic" by Gathercole & Pickering, 2001; see also Lecerf & de Ribaupierre, 2005, for a similar distinction). Despite the fact that the different domains seem to represent separate categories, the continuity model assumes that these categories are neither unitary nor totally independent. First, within each category, different aspects can be separated—for example, visual features such as shape and color (Ambrose at al., 2016). Second, a category or component of a category may be closer to another category or to one of its aspects. This is the case for the visuospatial categories, which appear more interrelated among themselves than with the verbal category (Pazzaglia & Cornoldi, 1999), or for some spatial-simultaneous patterns, which assume visual properties (Logie & Pearson, 1997).

The vertical continuum, on the other hand, defines a WM task as a function of the degree to which the need for controlled or otherwise effortful processes is operationalized in terms of the number and complexity of manipulations that must be carried out on stored information. According to Cornoldi and Vecchi's (2003) model, the classical distinction between passive and active tasks (or simple and complex spans, short-term tasks and WM tasks, or similar distinctions) refers to the extreme poles of the continuum, but many cases of WM requests represent intermediate situations that may be located along different points of the continuum.

In the framework of the Cornoldi and Vecchi's (2003) model, each WM task can thus be described on the basis of these two horizontal and vertical continua. The model predicts that the interaction between the two may vary, however, depending on the resources and peculiarities of a given individual (or category of individuals). Factors that play an important part in this sense might be age and individual differences. In people who are weak in general or in a certain domain, a task generally defined as a passive WM task (i.e., the information only needs to be retained) might also demand active control resources because of their particular weaknesses. Evidence in favor of this hypothesis also comes from children with atypical development (for a review, see Cornoldi & Vecchi, 2003). To give an example of such a situation, Lanfranchi et al. (2004) reported that individuals with Down syndrome differ from children with typical development in verbal WM tasks, regardless of how much attentional control the task requires. Individuals with Down syndrome typically have verbal abilities below their mental age, so the operations required in a verbal WM task may exceed their resources regardless of the active control demanded by the task. However, their visuospatial resources are in line with their mental ages. Lanfranchi et al. (2004) reported that in the case of visuospatial WM, the differences between typically developing children and individuals with Down syndrome varied depending on the levels of cognitive control required. There were no differences in a simple visuospatial span task, but greater differences in the increasing levels of cognitive control each task required. In a similar vein, Carretti et al. (2010) tested the hypothesis that the vertical continuum may appropriately describe the WM weaknesses of people with intellectual disabilities. Using five tasks assumed to vary along the vertical continuum from a very low-control request to a very high-control request (forward span, backward span, selective span, span with a dual task request, and updating), they found that the difference between a group of intellectually disabled individuals and controls increased linearly along with increases in the required control level.

In short, although various WM structure models have been proposed, the evidence produced to support them comes mainly from research on older children and adults. Largely because of the difficulties of devising age-appropriate assessment procedures, evidence concerning young children (and preschoolers especially) is scarce. Considering that younger children may present a distinct WM structure, this is a weakness in WM research.

Consequently, the main aim of our study was to investigate WM architecture in a large sample of children from 3 to 8 years old, using a large battery of age-appropriate tasks. Understanding the WM structure in young children is of fundamental importance because WM is crucial to various aspects of learning (Alloway, 2009), including reading decoding and comprehension (Christopher et al., 2012) and mathematics (Bull & Scerif, 2001). In fact, WM is an area of weakness in several neurodevelopmental disorders, such as autism (Giofrè et al., 2019), various learning disabilities (Swanson, 2020; Toffalini et al., 2017), and intellectual disability (Lanfranchi, 2013).

## How Working Memory Develops

Some attempts have been made to elucidate the structure and developmental stages of WM in typically developing children. Several independent studies have claimed that the classical tripartite model Baddeley and Hitch (1974) proposed seems to fit the data better than other models in school-age children (e.g., Alloway et al., 2006; Bayliss et al., 2003). A comparison of the findings of these studies revealed some differences in the strength of the association between the components of the tripartite model, however, depending on the children's age.

In one of the first studies among children from 4 to 15 years old (Gathercole et al., 2004), Baddeley and Hitch's (1974) classic three-factor model fitted the data better than a two-factor model (verbal vs. visuospatial), but only from 6 years of age onward. Neither model was tested on 4- or 5-year-olds, however, because the tasks tapping the CE were unsuitable for such young children. In a later study, Alloway et al. (2006) adapted the tasks used by Gathercole et al. (2004) and tested children from 4 to 11 years old. Here again, the tripartite model proved superior to several others. The authors found visuospatial STM and the CE to be very closely related. The correlation between the CE and the visuospatial sketchpad was .97 in children 4–6 years old, .82 in those aged 7–9, and .71 in those aged 9–11. Michalczyk et al. (2013) corroborated these early findings, demonstrating a strong correlation between the CE and the visuospatial sketchpad in children between 5 and 12 years old and finding no evidence of any drop in latent factor correlations from younger to older children. Campos et al. (2013) also found the correlation between these two components extremely high in children 8–9 years old ($r=.91$), high enough to support a two-factor model instead of a three-factor one.

In short, these results seem to indicate that visuospatial STM tasks are highly demanding for the CE, particularly in younger children (Alloway et al., 2006; Pickering, 2001). This finding is generally interpreted in line with the evidence that younger children use controlled attention when performing visuospatial tasks (see Cowan, 2005). However, other studies (e.g., Gathercole et al., 2004; Michalczyk et al., 2013) have found a strong correlation between the phonological loop and the CE.

Some studies have considered alternative models, such as Cowan's (2005) proposal of a domain-general attentional component common to different WM tasks (e.g., Gray et al., 2017; Hornung et al., 2011) or a distinction between a common storage and a common nonstorage component (Hornung et al., 2011). When Hornung et al. (2011) examined the WM structure in 5- to 7-year-olds, they administered a series of tasks measuring verbal STM and verbal and visuospatial WM. Their results did not support the classical Baddeley and Hitch (1974) model. Instead, the best fitting models (e.g. Cowan, 2008; Engle et al., 1999) suggested that both simple- and complex-span tasks tap the WM capacity in children. In fact, the WM structure included both a common storage component and a common nonstorage component.

## Aims and Hypotheses

Our main aim was to further examine the WM structure in children from 3 to 8 years of age. After analyzing age differences in each WM task, we tested a series of alternative models featuring modality-specific factors and including or excluding a general factor representing the CE (see Figure 1). The CE was featured using bifactor models, because we argued that all the tasks should partly involve attentional control in addition to the characteristics of the information to be processed.

In particular, Model 1 is a unitary model with only the general factor. Model 2 distinguishes between a verbal and a visuospatial component of WM, as in the tripartite model (Baddeley & Hitch, 1974), but without CE (Shah & Miyake, 1996). Model 3 further distinguishes between visual and spatial components in the visuospatial sketchpad (Logie, 1995), envisaging a three-factor structure based on the stimulus modality (verbal, visual, or spatial). Model 4 further distinguishes between spatial-sequential and spatial-simultaneous (or dynamic and static) components (Alloway et al., 2006; Pickering, 2001). In Model 4, we identified four factors: verbal, visual, spatial-sequential, and spatial-simultaneous (see Pazzaglia & Cornoldi, 1999). Finally, Models 5–7 are bifactor variations of Models 2–4, including orthogonal specific factors plus a domain-general component affecting all memory span tasks that can be interpreted either as general short-term storage (Colom et al., 2006) or as a CE factor. Its interpretation depends on its loading: If the loadings of the CE factor on the high-control tasks are consistently higher than those on the other tasks, the domain-general component can be interpreted as reflecting the CE.

In an additional analysis, we further refined our analysis of the CE component using a modified bifactor model in which the CE factor loaded on high-control tasks only, and specific components were allowed to load on their respective factors. This approach is unusual for a bifactorial model, but reasonable if the aim is to test the loading of high-control tasks on a domain-general component (i.e., a CE factor). The CE can be interpreted in this way.

We tested the WM model revealing the best fit among children of two different age groups in preschool (aged 36–67 months) and primary school (aged 68–100 months) and compared it between the two. Previous evidence has suggested that younger children may still have WM cognitive difficulties, whereas older children are better able to cope with the greater demands of WM tasks (Gathercole et al., 2004). If so, we would expect tasks requiring higher levels of cognitive control to load their respective factors more. Consequently, we used factor loadings to identify any common attentional factors (as Cowan's, 2005, model suggests). We also examined the correlations between factors to see whether the factors were distinctive—in other words, to ensure the correlations were not so high that they could make the different factors empirically indistinguishable.

In agreement with the literature, we expected to find increased performance in all the tasks across ages. Concerning the main objective of the paper, previous studies on children of the same age have suggested the existence of both domain-general and domain-specific components. In line with this body of research (e.g., Hornung et al., 2011), we hypothesized that the WM structure can be differentiated at an early age and further subdivided in the visuospatial component (Logie & Pearson, 1997). Our models also tested the presence of a CE factor. As already discussed, previous findings have varied regarding the presence of a CE factor in very young children (Gathercole & Pickering, 2001). This sometimes depends on the characteristics of the task or the statistical approach used to derive this domain-general factor.

To overcome these issues, we decided to use a battery of WM tasks devised in previous research (Belacchi et al., 2010; Carretti et al., 2010; Lanfranchi et al., 2004) in which domain-general and domain-specific aspects were varied. For each domain-specific WM component, the battery of tasks included not only a low and a high control but also a medium-control task, allowing us to examine domain-general components. These tasks were devised in an attempt to equalize the overall difficulty of each task while simultaneously increasing control demands. For these characteristics, the tasks were appropriate for testing very young children, avoiding problems in deriving a CE component due to task difficulty.

## Method

**Participants**

Our sample consisted of 739 children (51% males) attending preschool or primary school in northeastern Italy, mainly of European origin (84%). The remainder of the students were of African (9%) or Asian origin (7%). All students spoke Italian as their first language and came from medium to medium-high socioeconomic backgrounds. Participants ranged from 36 to 100 months old, $M_{age} = 64.3$ months ($SD = 13.13$). For the purposes of a multigroup confirmatory factor analysis (CFA) based on age, the children were divided into two groups. The younger group included 501 children between 36 and 67 months old, $M_{age} = 56.8$ months ($SD = 6.4$, 48% males), and the older group included 238 children between 68 and 100 months old, $M_{age} = 80.0$ months ($SD = 9.0$, 58% males). Although this distinction led to unequal group sizes, it allowed us to distinguish between preschoolers and primary school children. The children in the two groups all attended mainstream preschools or primary schools. Parental consent was obtained for all children participating in the study.

It is worth noting that in the Italian school system, children from 6 years of age onward are typically enrolled in primary schools, whereas children below 6 years of age generally attend preschools. Therefore, in the Italian school system, the division between children enrolled in preschools and primary schools is natural. In fact, children in preschools typically receive informal teaching focused on games and other activities. In primary school, teaching becomes more formal, and children start to learn reading, writing, mathematics, and other subjects.

**Materials**

We administered a battery of 12 WM tasks—three for each stimulus presentation modality (verbal, visual, spatial-sequential, and spatial-simultaneous). For each domain, the tasks involved a gradually increasing level of active control over the information during processing and maintenance, as Cornoldi and Vecchi (2003) suggested. We manipulated the level of control involved in each task according to the number of mental operations to be performed at the same time. For each presentation modality, the material was the same, but three different tasks were administered. One involved attentional resources to a lesser extent, only requiring the retrieval of the material as presented. Another involved a moderate degree of manipulation, requiring selective recall of only some of the information presented. The third task demanded a high level of cognitive control, requiring the selection and retrieval of specific items and the performance of a concurrent task. To avoid a floor effect for the more complex tasks, a constant difficulty was maintained across the tasks, reducing the amount of information to be remembered as the control requirement increased.

We used a self-terminating procedure for all tasks. Participants were presented with a series of increasingly complex pairs of trials in each task, with two trials each for four levels of difficulty. The task continued as long as a child was able to solve at least one of the two trials on a given difficulty level. The tasks were adapted from previous studies (e.g., Lanfranchi et al., 2015) and had good psychometric properties. For each task, participants first completed some practice trials at the lowest level of difficulty, and the task was administered only when the child appeared to understand what it entailed. For the tasks requiring medium or high degrees of cognitive control, the instructions and practice trials were carefully designed to make the child feel confident about the task and to ensure that all of the material was processed, even though only part of it had to be recalled. An absolute credit score was used to facilitate comparison between different tasks. We calculated overall scores as the sum of the trials completed perfectly (min = 0, max = 8) for each task. The same set of tasks was used in both sample groups (see Table 1).

Table 1 about here

**Verbal Tasks**

For the verbal tasks, the material consisted of a series of concrete and familiar two-syllable words, as used in previous studies (e.g., Lanfranchi et al., 2004).

*Verbal, Low Cognitive Control*

Participants were presented with an oral list of words and asked to recall the words immediately afterward in their order of presentation. The lists gradually became longer, from a minimum of two words to a maximum of five. A score of 1 was awarded if a list was recalled accurately.

*Verbal, Medium Cognitive Control*

Participants were presented orally with one or two lists of words and told they would be asked afterward to repeat the first word from each list. There were four levels of difficulty: In the first two trials, children were required to recall the first word in a list of two or three words; the third and fourth trials, however, consisted of two lists, and children had to remember the first item of each.

To help participants understand the concept of a series, the orally presented words were written on colored cards, with all words on the same list written on the same card and each new list written on a different-colored card. The tester read the words from the cards, and the children could see the color of the card but not the words written on it. After reading each word list, the tester would put the card or cards on the table with the side containing the written words face down and then ask the child to name the first word from the red card and the first word from the blue one, for example. A score of 1 was awarded for every trial performed correctly (when two lists were used, the trial was only performed correctly if the first word on both lists was recalled).

*Verbal, High Cognitive Control*

Participants were presented orally with a list of two to five words. They were asked to remember the first word on the list and to tap on the table when they heard the word *palla* (ball). The word "ball" was present in each trial. In this case, tapping served as a secondary task. A score of 1 was awarded for every trial completed correctly (i.e., when a participant remembered the first word of the list and tapped the table at the right moment).

**Visual Tasks**

The material for the visual tasks was drawn from Cornoldi and Vecchi (2003) and consisted of a series of stimuli representing houses of different shapes.

*Visual, Low Cognitive Control*

Participants were given 8 s to look at a series of houses drawn on a piece of cardboard. Then the cardboard was removed, and they were given several cards, each showing one of the houses seen before. The child was asked to place them in their previously seen positions. The task involved four levels of difficulty depending on how many houses had to be positioned (two, three, four, or five). A score of 1 was awarded for each trial completed correctly.

*Visual, Medium Cognitive Control*

Participants were given 8 s to look at a series of houses drawn on a piece of cardboard. Half were colored red, and half green. Then the cardboard was removed, and they were given several cards, each showing one of the houses in black and white. The child was asked to position only the houses that had previously appeared in a given color (only the red ones in one trial and only the green ones in another). There were four levels of difficulty depending on the number of houses presented (three, four, or five) and the number of houses to be remembered (two or three). A score of 1 was awarded for each trial performed correctly.

*Visual, High Cognitive Control*

Participants were given 8 s to look at a series of houses drawn on a piece of cardboard. Half of them were colored red; in half of the trials, one of the other houses was colored blue. Participants were required to recall red houses. When the blue house appeared in the series, participants also had to tap their hand on the table (concurrent task). Although the children only had to tap their hands on the table in half of the trials, the concurrent task required increased attention in all trials (i.e., children needed to scan all the images to detect the blue house). This task subtracted resources from the primary memory task and for this reason served as a concurrent task. A score of 1 was awarded only when the child performed both the sorting task and the concurrent task correctly.

**Spatial-Simultaneous Tasks**

For all the spatial-simultaneous tasks, the material consisted of $2 \times 2$, $3 \times 3$, or $4 \times 4$ matrices in which some squares could be colored in; the tasks were adapted from previous studies (e.g., Lanfranchi et al., 2009).

*Spatial-Simultaneous, Low Cognitive Control*

Participants were given 8 s to look at the positions of some green squares in a matrix. Then the matrix was removed, and they were asked to remember the locations of the green squares and point to their positions on a blank matrix. The task involved four levels of difficulty depending on the number of squares to remember (two or three) and the size of the matrix, which was $2 \times 2$ on the first level, $3 \times 3$ on the second and third levels (with two and three green squares, respectively), and $4 \times 4$ on the fourth level (with three green squares). There were two trials for each level of difficulty. A score of 1 was awarded for every matrix recalled completely correctly.

*Spatial-Simultaneous, Medium Cognitive Control*

The procedure was the same as in the previous task, but in this case, the colored squares could be red or green, and the children had to remember only the positions of the squares in a given color. After the matrix was removed, participants were shown a blank matrix and asked to point out the locations of the red (in half the trials) or green squares (in the other half). They were not warned which color they would need to recall. The task had five levels of complexity depending on the number of squares to recall (from two to three) and the size of the matrix, which was $2 \times 2$ on the first level, $3 \times 3$ on the second and third levels (with two or three green or red squares, respectively), and $4 \times 4$ on the fourth level (with two or three green or red squares). There were two trials for each level of difficulty. A score of 1 was awarded for every matrix recalled completely correctly.

*Spatial-Simultaneous, High Cognitive Control*

The procedure was the same as in the previous tasks, but the squares to remember were all colored red; the matrix sometimes also contained a blue square (in half of the trials), in which case participants had to tap on the table with their hand. Then they had to point at the positions of the red squares on an empty matrix. The task included four levels of difficulty depending on the number of squares to recall (two or three) and the size of the matrix, which was $2 \times 2$ on the first level, $3 \times 3$ on the second and third levels (which contained two and three red squares, respectively), and $4 \times 4$ on the fourth level (with three red squares). To carry out the tapping task accurately, the children had to maintain their attention to detect the blue square. This activity subtracted attentional resources from the main task and for this reason was considered concurrent task. A score of 1 was awarded when the child performed both the recall and the concurrent tapping task correctly. In every other case, they scored 0.

**Spatial-Sequential Tasks**

The stimuli for the spatial-sequential tasks comprised matrices and paths within them; the tasks were adapted from previous studies (e.g., Lanfranchi et al., 2009).

*Spatial-Sequential, Low Cognitive Control*

Participants were shown the path taken by a small frog on a $3 \times 3$ or $4 \times 4$ matrix and immediately afterwards were asked to repeat the frog's steps from cell to cell. There were four levels of difficulty depending on the number of steps along the frog's path and the dimensions of the matrix ($3 \times 3$ for the first level of difficulty, with the frog taking two steps; $4 \times 4$ for the other levels, with the frog taking two, three, and then four steps). The frog's steps were presented at a rate of approximately one step every 2 s. A score of 1 was awarded for every path recalled correctly.

*Spatial-Sequential, Medium Cognitive Control*

As in the previous task, participants were shown one or two paths taken by the frog on a 4 × 4 matrix, and they were informed that their task was to remember the frog's starting position(s). The level of complexity depended on the number of paths and the number of steps along each path. The first and second levels involved one pathway with two and three steps, respectively. For the third and fourth levels of difficulty, there were two paths with three and four steps, respectively. Accordingly, in the first four trials, the child had to remember only one path; in the fourth to eighth trials, the child had to remember two positions per trial. A score of 1 was awarded for every trial completed correctly.

*Spatial-Sequential Task, High Cognitive Control*

Participants were asked to remember the frog's starting point along a path on a 4 × 4 matrix (where one of the 16 cells was colored red) and to tap on the table whenever the frog jumped onto the red square. The frog jumped onto the red square once in each trial. The position from where it jumped onto the red square varied across trials. The task included four levels of difficulty depending on the number of steps in the path (from two to five). A score of 1 was awarded for every trial completed correctly (i.e., when a participant remembered the path's starting point and tapped the table at the right moment). In all other cases, they scored 0.

**Procedure**

Testing was done in a quiet, well-lit room. Participants were individually assessed during two sessions lasting approximately 30 min each. The order of presentation of the tasks associated with each modality (verbal vs. visual vs. spatial-simultaneous vs. spatial-successive) was counterbalanced across participants. However, based on previous evidence, we preferred to maintain the same order within each modality, starting with the low-control task implying the simplest request and then moving to the medium- and high-control tasks. Therefore, within each modality, the order was always the same, counterbalanced between modalities. The study was approved by local Ethics Committee of the School of Psychology - University of Padova (title of the project "Working memory and learning" protocol number 3663).

**Data Analysis Strategy**

A Bayesian approach to data analysis was adopted because it presents several advantages (for a review, see Vandekerckhove et al., 2018). Bayesian modeling emphasizes the estimation of model parameters and uncertainty, allowing for a more nuanced, probabilistic description of the phenomenon of interest instead of reducing it to a series of accept/reject decisions (e.g., Kruschke & Liddell, 2018; McElreath, 2016). A Bayesian approach also allows the handling of complex statistical models (e.g., Vandekerckhove et al., 2018) and facilitates convergence in the presence of very sophisticated, complex models.

In a preliminary analysis, we visually examined the univariate and bivariate distributions of all the variables (see Figure S1 and Table S1 in the supplemental online material), and we examined age trends (see Figure S2 in the supplemental online material). Then we used a CFA to fit and compare the seven models obtained, corresponding to different hypotheses on the WM structure (Figure 1). Each model was fitted using the Bayesian Markov chain Monte Carlo estimation method implemented in the Stan programming language (Stan Development Team, 2017) coupled with the R-package blavaan (Merkle & Rosseel, 2016), and uninformed default priors were used for the model parameters.[1]

The overall goodness of fit was examined using several criteria. For instance, the convergence criterion (CC) is calculated as the percentage of converging parameters in a model. For each parameter estimated, convergence was determined from the potential scale reduction statistic $\hat{R}$, which measures the ratio of the average variance of samples within each chain to the variance of the pooled samples across chains; if all chains are at equilibrium, $\hat{R} = 1$ (Gelman & Rubin, 1992). The CC represents the percentage of parameters for which $\hat{R}$ is below the cutoff value of 1.1. A CC of 100 means that all model parameters reached convergence.

Another set of criteria comprised Bayesian versions of classical CFA fit indices (Jorgensen & Garnier-Villarreal, 2018). We calculated the Bayesian comparative fit index, BCFI [0, 1], with large being good; the Bayesian Tucker-Lewis fit index, BTLI [0, 1], with large being good; and the Bayesian root mean square error of approximation, BRMSEA ≥ 0, with small being good. According to common rules of thumb, a BCFI and BTLI > .90 and a BRMSEA < .10 are minimum requirements for a fit, and a BCFI and BTLI around .95 and a BRMSEA around or below .05 are optimal. The widely applicable information criterion (WAIC) is calculated by taking the averages of log likelihood over the posterior distribution. As in other commonly used information criteria (e.g., Akaike information criterion), smaller values in the WAIC indicate a better fit (McElreath, 2016; Vehtari et al., 2017).[2] This index can be used to derive Akaike weights, which indicate the relative probability of each model being the best of a set, given the data (Burnham & Anderson, 2002; McElreath, 2016; Wagenmakers & Farrell, 2004). For the CFAs, each model was fitted using the Bayesian Markov chain Monte Carlo (MCMC) estimation method. We sampled the posterior distributions of parameters by running four MCMC estimations with 1,000 replicates each. For each chain, the first 500 replicates were burning iterations, so posterior distributions were based on a total effective sample of 2,000 replicates.

First, we conducted a series of CFAs separately by age group to establish the best model for each group. We simultaneously ran multigroup CFAs to establish whether the best fitting model at the multigroup level was the same as the best fitting model for each age group. Second, we examined the parameters of the best fitting multigroup model in detail. The mean values of the posterior distributions were used as the parameter estimates. As measures of uncertainty, we calculated the 90% highest posterior density intervals (HPDIs), which are similar to 90% confidence intervals in the frequentist framework. We did not try to establish statistical significance for any parameter, however. Instead, following the traditional Bayesian approach, we aimed to identify credible estimates with their uncertainty.

**Invariance Assessment**

We assessed invariance between groups as follows. First, if the best fitting model was the same for both age groups and at the multigroup level, configural invariance was established. Then we examined further steps of model invariance by considering how the posterior distributions of the model parameters differed between the groups. In the traditional invariance approach, degrees of invariance are investigated by constraining parameters to be equal across groups (e.g., Beaujean et al., 2012). Using the Bayesian approach, parameters are always considered random variables rather than fixed quantities (see, e.g., Muthén & Asparouhov, 2012). Therefore, we did not impose any further model constraints. All parameters were freely estimated in the two groups, and we assessed the degrees of invariance by computing the degrees of overlap in the posterior distributions of the estimated parameters across the groups. Specifically, we calculated the posterior distributions of the between-group differences in loadings (to assess metric invariance) and intercepts (to assess scalar invariance) and their 90% HPDIs. We also estimated the percentage of overlap between the posterior densities (using the Overlapping R package; Pastore, 2018). The overlapping percentage ranges from 0 (when the posterior distributions are completely separate) to 100 (when they coincide perfectly).

This study was not preregistered. The dataset has been made publicly available at the Open Science Framework (OSF) repository and can be accessed at https://doi.org/10.17605/OSF.IO/EP46D (Toffalini, 2022).

**Results**

The univariate distributions of all variables and descriptive statistics and correlations are reported in the supplemental online material (Figure S1 and Table S1, respectively). The age trends estimated with linear models showed lower intercepts but larger slopes for high-control than for low-control tasks; for medium-control tasks, the intercepts were generally similar to those of the low-control tasks, but the slopes were similar to those of the high-control tasks, with a clearer pattern for verbal than for spatial and especially visual tasks (see supplemental online material, Figure S2). Younger children usually achieved a lower performance in attention-demanding tasks (medium- and high-control tasks) than older children; changes in performance across ages were generally steeper for medium- and high-control tasks than for low-control tasks.

*Figure 1 about here*

**Confirmatory Factor Analysis and Model Comparisons**

Figure 1 shows the seven models that were fitted, which we described in the introduction. Table 2 shows the fit indices for the seven models fitted with a multigroup CFA and for each of the two age groups. All models converged (all CCs = 100). Model 4 had acceptable fit indices in all cases. The three bifactor models had acceptable fit indices, but their WAICs were much higher than those of Model 4, which the Akaike weights largely favored in the multigroup CFAs and in each of the two groups (the Akaike weight for Model 4 always exceeded 90%). Regarding the multigroup models, for Model 4, the results were BCFI = .95, 90% HPDI (.94, .96), BTLI = .93, 90% HPDI (.92, .94), BRMSEA = .07, and 90% HPDI (.06, .07). Based on these findings, we concluded that Model 4 had the best fit, and it was therefore retained as our final model. Based on Model 4, the specific factors comprising WM in young children according to our battery of tests refer to the type of material—verbal versus visual and spatial-simultaneous versus spatial-sequential. These appear to be sufficient because they do not require an additional general factor associated with general STM storage or executive control.

*Table 2 about here*

**Comparisons Between Age Groups and Invariance**

Because the same Model 4 emerged with the best fit in the multigroup CFA and when each of the two age groups were analyzed individually, with good fit indices in all cases, we concluded with a multigroup configural invariance. For the other invariance steps, we moved on to consider the model parameters. For all subsequent analyses, we examined the posterior distributions of the parameters from the multigroup CFA model.

Concerning the loadings, visual inspection showed a general between-group overlap across tasks (Figure 3). The average overlap between the posterior distributions in the two groups was 31%. All between-group differences in estimates were small, however, with none exceeding .20 and most below .10 in absolute value. No systematic tendencies emerged in the differences across the two groups. This means that there is a relatively good degree of metric (weak) invariance between the two groups in the overall factorial model, although some between-group divergences emerge locally in the loadings. Further details can be found in the supplemental online material.

Concerning the intercepts, it clearly emerged that they were higher in older than in younger children. This was true for all 12 tasks. It was only in two tasks (i.e., visual with low and high control) that the between-group difference in intercepts had 90% HPDIs failing to exclude zero. The average estimated overlap between the posterior distributions of intercepts in the two groups was only 5%. This means that there is no strong (scalar) invariance between the two groups in the factorial model. Further details can be found in the supplemental online material.

When we looked at factor correlations, the average overlap between the posterior distributions of the two groups was a modest 19%. However, most of the differences between the correlations in the two groups were practically negligible, below .12. The only two correlations that differed substantially were the one between the verbal and spatial-simultaneous factors, which was stronger in older than in younger children, $\Delta = .28$, 90% HPDI (.15, .39), and the one between the visual and spatial-sequential factors, which was stronger in younger than in older children, $\Delta = -.58$, 90% HPDI ($-.74$, $-.43$). All factor correlations are shown in Figure 2. Further details can be found in the supplemental online material.

The estimated standardized residual variances also are shown in Figure 2. The differences between the two groups were modest and obviously reflected the differences in the standardized loadings, so they are not discussed further. All details can be found in the supplemental online material.

*Figure 3 about here*

**Cronbach's Alpha**

We calculated Cronbach's alpha for the factors emerging from the best-fitting model (Model 4) by age group. The standardized Cronbach's alpha calculated from the posterior distributions of the factor correlations can be interpreted as a measure of the factors' reliability. The value was good and similar for both groups but slightly better for the younger than for the older children ($\alpha = .85$ for the former, and $\alpha = .80$ for the latter).

When we computed Cronbach's alpha for each factor emerging from Model 4, the alphas were calculated on the correlations observed between the tasks. For the younger group, the reliability was slightly below the conventional threshold for the verbal factor ($\alpha = .65$), acceptable for the visual factor ($\alpha = .70$), and good for the spatial-simultaneous ($\alpha = .79$) and spatial-sequential ($\alpha = .80$) factors. The older group had generally similar values (verbal, $\alpha = .61$; visual, $\alpha = .72$; spatial-simultaneous, $\alpha = .72$; and spatial-sequential, $\alpha = .75$).

**Additional Analyses**

A bifactor model formalizes the idea that the performance in each task simultaneously reflects a domain-specific factor (e.g., verbal, visual) and a domain-general factor (e.g., the CE). In our case, the bifactor models failed to prevail over Model 4, which features only the specific factors. This suggests that a domain-general factor has a limited relevance in our data. Nonetheless, it is still possible that a domain-general factor could emerge if it was specified differently. In particular, bifactor models generally require the factors to be orthogonal to converge, but this may negatively affect the fit. Therefore, we considered an additional analysis that was plausible based on our results, including four correlated specific factors (as Model 4), plus an additional uncorrelated CE factor that loads only on the high-control task. This CE factor should model the residual variance that is left between the high-control tasks if they reflected, at least in part, a CE. We avoided also loading it on the medium-control tasks both to avoid convergence problems and because most of the correlated residual variance is theoretically expected to remain from the high-control tasks. The model, which we labelled Model 4CE, is depicted in Figure 4.

In younger children, Model 4CE had a good fit: BCFI = 0.95, BTLI = 0.92, BRMSEA = 0.07, WAIC = 22,864, and it slightly outperformed the original Model 4, $\Delta$WAIC = −2 (in the younger group; an equality constraint had to be imposed on the CE loadings to facilitate convergence). In older children, Model 4CE also had a good fit: BCFI = 0.96, BTLI = 0.93, BRMSEA = 0.06, WAIC = 10,648, and it performed just as well as the original Model 4, $\Delta$WAIC = 0. This analysis suggests that a CE factor can be included in the already well-fitting Model 4, if adequately specified. Nonetheless, we should consider how much the 4CE Model adds in terms of the explained variance. Table 3 reports the standardized loadings and shows that the additional CE factor explains a small portion of the variance in the high-control tasks in both the younger and older children. Despite this, Model 4CE appears interesting because it offers the possibility of considering the existence of the CE.

Figure 4 and Table 3 here

Finally, we conducted an additional analysis on another model featuring no domain-specific factors but only three factors representing low, medium, and high control. The model had an inadequate fit. It is detailed in the supplemental online material in Figure S7 and the comments.

## Discussion

WM is a core component of human cognition, and understanding its structure in younger children is important, partly because of the associations between WM and other aspects of cognitive functioning in childhood and partly because the WM structure described in adults or older children may not necessarily apply to younger children (Alloway et al., 2006; Logie & Pearson, 1997; Pickering et al., 2001). Hence, our study was performed on the WM structure in children from 3 to 8 years old, an age group for which evidence is scarce and unclear, partly due to the difficulty of finding age-appropriate tasks. We tested several different WM models using a large battery of WM tasks, which varied in terms of how the information to process and recall was conveyed and the active control required.

Because our sample included participants of different ages, we analyzed cross-sectional trends first. Interestingly, the intercepts and slopes varied depending on the active control the task required: The intercepts were generally lower in the case of high-control tasks than in low-control tasks. In the case of medium-control tasks, intercepts were more similar to high-control than low-control tasks. In other words, younger children usually achieved lower performance in attention-demanding tasks (medium- and high-control tasks) than in low-control memory tasks that involved only maintenance. Moreover, the changes in performance across ages were generally steeper for medium- and high-control tasks than for low-control tasks. These results indirectly demonstrate that the manipulation of cognitive control was effective: Age differences are greater in tasks that require managing more than one mental operation to perform (as required in the medium- and high-control tasks).

Regarding the main aim of the paper—the analysis of the WM structure in a sample of children 3–8 years old—we tested different models: one considering WM as completely unitary (Model 1), three emphasizing the role of domain-specific components (Models 2–4), and the rest testing the existence of a general factor and capturing the common characteristics of the WM measures (Models 5–7). A four-factor model, involving only the domain-specific components of WM, was the best-fitting model, suggesting that the WM structure is differentiated according to the domains in very young children. We also tested the invariance of the final WM model separately in the two groups—younger children 3–5 years of age and older children 6–8 years of age. Older children tended to have better overall performances than younger children, as demonstrated by the difference in the intercepts (discussed above), whereas the structure of the two groups' WM seemed to be the same (Figure 2). This invariance also concerns the internal relations within each factor. We examined this aspect qualitatively, analyzing factor loadings—for which, with few exceptions, the overlap tended to be homogeneous across the groups (see Figure 3). The strength of the correlations between factors, with some exceptions, was similar for the two groups as well. Our results generally identified a high, but not extremely high interfactor correlation, indicating that the factors included in the final model were distinguishable in both groups of children.

The best-fitting model did not include an executive factor, however. To understand this result better, in an additional analysis, we reformulated the model without imposing orthogonal constraints between latent factor correlations, except for the correlation between the CE factor and the other factors. Notably, although the orthogonal constraint is theoretically plausible, particularly in intelligence research in which the g factor is the main aim of the analyses, this might not be entirely reasonable in WM research, in which an executive component probably has different peculiarities than the g-factor. We allowed this executive factor to load only on high cognitive control tasks. The inclusion of this factor, which was based on previous literature (e.g., Hornung et al., 2011), showed a modest improvement of fit (or equal fit in older children). In addition, looking at the variance explained (see Table 3) the contribution to the explained variance of high-control tasks on the CE factor was generally much lower than the contribution of the same tasks to the domain-specific factors.

On one hand, our findings support a multicomponent view of WM with independent modality-specific components (as suggested by Shah & Miyake, 1996, and partly by Cornoldi & Vecchi, 2003). In comparing our findings to the classical tripartite model Baddeley and Hitch (1974) proposed, we also found a further articulation of the visuospatial component, with the visual material being distinguishable from the spatial material (e.g., Logie, 1995) also in very young children. Interestingly, the distinction within the spatial component, depending on the presentation format, was also evident, as has been demonstrated in adults and older adults.

On the other hand, the additional analyses suggested that the four high-control tasks share a domain-general component: Although the CE factor improved the model fit (only in younger children), the increase in explained variance was less than modest. High-control tasks contributed to a greater extent to the domain-specific components than to the domain-general factor (see Table 3).

In connecting these results to the previous literature on WM and executive functions, the prevalence of a model based on domain-specific factors over a model including a domain-general factor appears odd. Several studies have suggested that executive function is undifferentiated in early development (e.g., Wiebe et al., 2011; Willoughby et al., 2012) and that WM is a differentiated component with respect to executive functions after age 6 years old (e.g., Lee et al., 2013), with other evidence indicating that it separates after age 9 (Shing et al., 2010). All these studies, however, considered WM a unique construct and did not include further analysis of its structure: Therefore, a full comparison of the current and previous findings is difficult.

It is also possible that the preference for a domain-specific model may depend on the overrepresentation of tasks assessing visuospatial processing. Some evidence has suggested that in young children, the executive factor is strongly related to and almost indistinguishable from the visuospatial factor (Campos et al., 2013; Michalczyk et al., 2013), and this is particularly true for younger children (Alloway et al., 2006). It is worth noting, however, that compared to the aforementioned studies, we used a different approach in which the executive component is obtained after partialling out the variance of the other factors. Therefore, our results are not directly comparable to evidence obtained using a different methodological approach.

The emergence of the CE factor would have been in line with the research on adults, where when statistically controlling for the variance in conjunction with low cognitive control tasks (referred to as STM tasks in part of the literature), researchers obtained a residual component corresponding to the CE (referred to as WM; Engle et al., 1999). For example, Hornung et al. (2011) used this approach in a sample of children from 5 to 7 years of age and found a residual WM variance (i.e., CE) after partialling out the variance of tasks with lower cognitive control. However, that study had a series of limitations; although it used a relatively large sample ($N = 161$), the number of indicators was not adequate for testing complex models (see Gray et al., 2017, on this point). In addition, having a very limited number of indicators for each factor means that a very large number of equality constraints have to be imposed to achieve convergence on complex models (Beaujean, 2014). This is no trivial issue, partly because imposing numerous constraints may well bias the results (e.g., Giofrè et al., 2019). Therefore, our work confirms and expands previous results using a different set of tasks, with an adequate sample size and with a sufficient number of indicators for each factor.

Our study generated some interesting findings but also some limitations to bear in mind during future research. Some of the limitations are related to the characteristics of the tasks we administered, especially in the case of the medium-control tasks. For example, one could argue that the requirement to ignore other items in the medium-load task does not necessarily add an additional control demand. It can also be argued that the medium-load manipulation is not comparable across modalities: For example, the medium-control spatial-simultaneous task could be more demanding than its verbal counterpart because it requires the maintenance of twice as many items in memory given that the tasks involve the selective, postcued recall of one color. Furthermore, the fact that the dual request was not present in all trials of the verbal and spatial-sequential high-control tasks might have made those tasks more difficult in some trials (those in which the concurrent task should have been carried out) than in others (where no concurrent task was required). However, we could not examine this aspect in our study due to the limited number of trials in each task.

The adoption of a self-terminating procedure might also have contributed to the pattern of results seen in our sample because participants encountered the most demanding level of difficulty for their capacity in each task. Our use of exactly the same material for the three different tasks involving the same stimulus modality may have emphasized their relationships also because the children could be influenced by practice effects. It is worth noting that the tasks for each WM component were generally selected because they were compatible with a hypothetical account/model of WM. For example, several studies were conducted traditionally with tasks designed to measure the very popular tripartite model or Baddeley's (2000) more recent model that also includes an episodic buffer and found evidence for models that were in some way in line with these accounts (Alloway et al., 2004, 2006; Bayliss et al., 2003; Giofrè et al., 2013). In fact, demonstrating a further distinction in visuospatial WM, for example, considering its visual versus spatial or simultaneous versus sequential components, is not possible without including tasks tapping into these specific factors. The same happened in our study: Omitting the tasks that assess the episodic buffer makes it impossible to detect. We treated the four WM components as categorical, without considering their overlaps or further specifications within each component; however, it is worth considering the possibility that some children use specific strategies that change some task characteristics. In fact, some nuanced differences between visual and spatial-simultaneous tasks at low, medium, and high control may reflect individual differences in the use of strategies that affect the tasks' difficulty.

Another limitation is that the results may reflect, in addition to the modality, the structural similarity of the tasks used to tap into the different components. This could also explain the high correlation between the verbal and sequential factors in general and for the older group in particular (as Mammarella & Cornoldi, 2005, suggested). In fact, in both verbal and sequential tasks, materials are presented sequentially, and this can explain the extremely high correlation between those two factors, which although highly correlated are still distinguishable, because about 30% of the variance is not shared among them.

Future studies are also needed to examine age changes across broad age ranges and using different tasks. With our sample, it was only possible to investigate age trends up to 8 years old, and the results indicated a clear age trend in virtually all tasks considered. It is also worth noting that depending on the age group, the correlation patterns between the verbal and spatial-simultaneous factors and between the visual and spatial-sequential factors were somewhat different. In the former case, the correlation increased and, in the latter, it decreased in younger and older children. This could suggest that the abilities underlying these factors unify or diversify with age. In that case, these patterns would not be completely consistent with the view of a common WM structure in both age groups. This issue should be addressed, for example, by expanding the age groups to include older children and particularly testing children from 11 years of age onward, when their WM structure might tend to be closer to that of adults.

Finally, in this study, we did not test whether different WM components, in particular our CE factor, could predict significant portions of the variance in other cognitive tasks such as reasoning, reading, or mathematics. Our results showed the possible presence of the CE factor, but with small loadings. Therefore, the inclusion in future studies of other variables for outcomes is particularly advisable for understanding the role of CE factor.

Apart from these considerations, this study has some strengths that we feel are worth emphasizing. In particular, our battery of tasks overcomes the common limits of WM tasks used in developmental research to date. Most complex span tasks suitable for adults or older children are inappropriate for young children, so the tasks used in the literature differed structurally from those generally used to analyze WM in other age groups. For example, Hornung et al. (2011) administered backward span tasks to measure WM executive processes, but backward span tasks involve different processes from those involved in classic complex span tasks (such as the reading span task or the operation span task). Such differences may affect the comparability of research findings on adults' and children's WMs. In contrast, we adapted the active tasks from classic WM tasks without altering the nature of the executive processes involved. The verbal, high-control task, for instance, was adapted from the categorization WM span task De Beni et al. (1998) devised to assess WM in young adults, which also has been used with children (Carretti et al., 2004) and individuals with neurodevelopmental disorders, such as children with ADHD (e.g., Re et al. , 2010) and individuals with intellectual disabilities (Lanfranchi et al., 2009, 2015).

Footnotes

[1] The models were expressed from the following equation: $Y = \tau + \Lambda\eta + \epsilon$

where: Y indicates the items observed, $\tau$ the intercepts, $\Lambda$ the factor loadings matrix, $\eta$ the factor scores, and $\epsilon$ the residuals. The following prior distributions were specified for the model parameters

$\tau_j \sim$ Normal (0, 32) [$\tau_j$ prior, $j = 1, …, 12$]

$\lambda_j \sim$ Normal (0, 10) [$\lambda_j$ prior, $j = 2, …, 12$]

$\theta_j \sim$ Gamma (1, 0.5) [$\theta_j$ prior, $j = 1, …, 12$]

$\phi_{ij} \sim$ Beta (1, 1) [$\phi_{ij}$ prior, $i, j = 1, …, 12$]

Intercepts ($\tau$) had normal priors with mean = 0 and SD = 32; factor loadings ($\lambda$) had normal priors with mean = 0 and SD = 10; residual variances ($\theta$) had Gamma (1, 0.5) priors, and the factor covariance matrix had Beta priors with $\alpha = 1$, $\beta = 1$.

[2] For *BCFI, BTLI*, and *BRMSEA* we also computed the 90% highest posterior density interval (HPDI), the narrowest interval containing the specified probability mass or, in other words, the interval containing the parameter values with the highest posterior probability (McElreath, 2016).

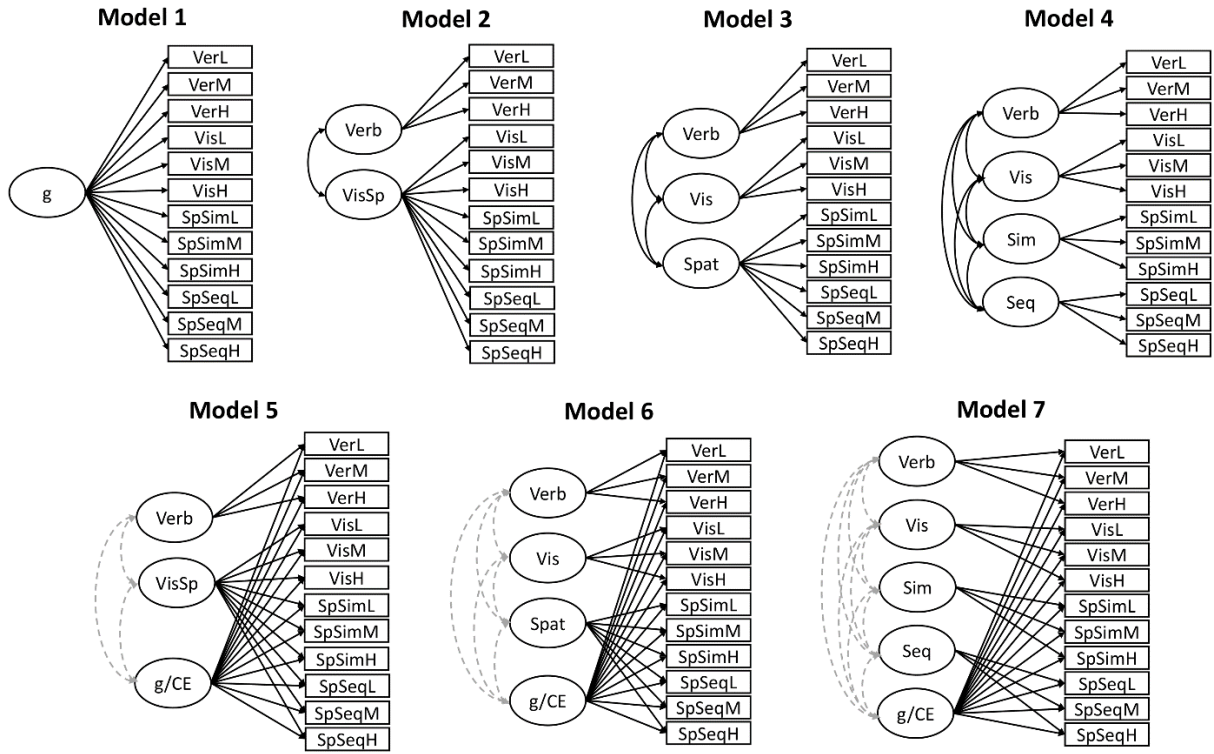

**Figure 1**. Graphical representation of the seven models considered. CE = central executive; G = *g* factor; G/CE = either *g* factor or central executive in Model 7, depending on the theoretical interpretation of the loadings; Verb or Ver = verbal; Vis = visual; Seq = sequential; SpSeq = spatial-sequential; Spat = factor; Sim = simultaneous; SpSim = spatial-simultaneous. The last letters of the names of the observed variables indicate the level of executive control: L = low, M = medium, H = high. The dashed gray correlations are set to zero, as the models are orthogonal.

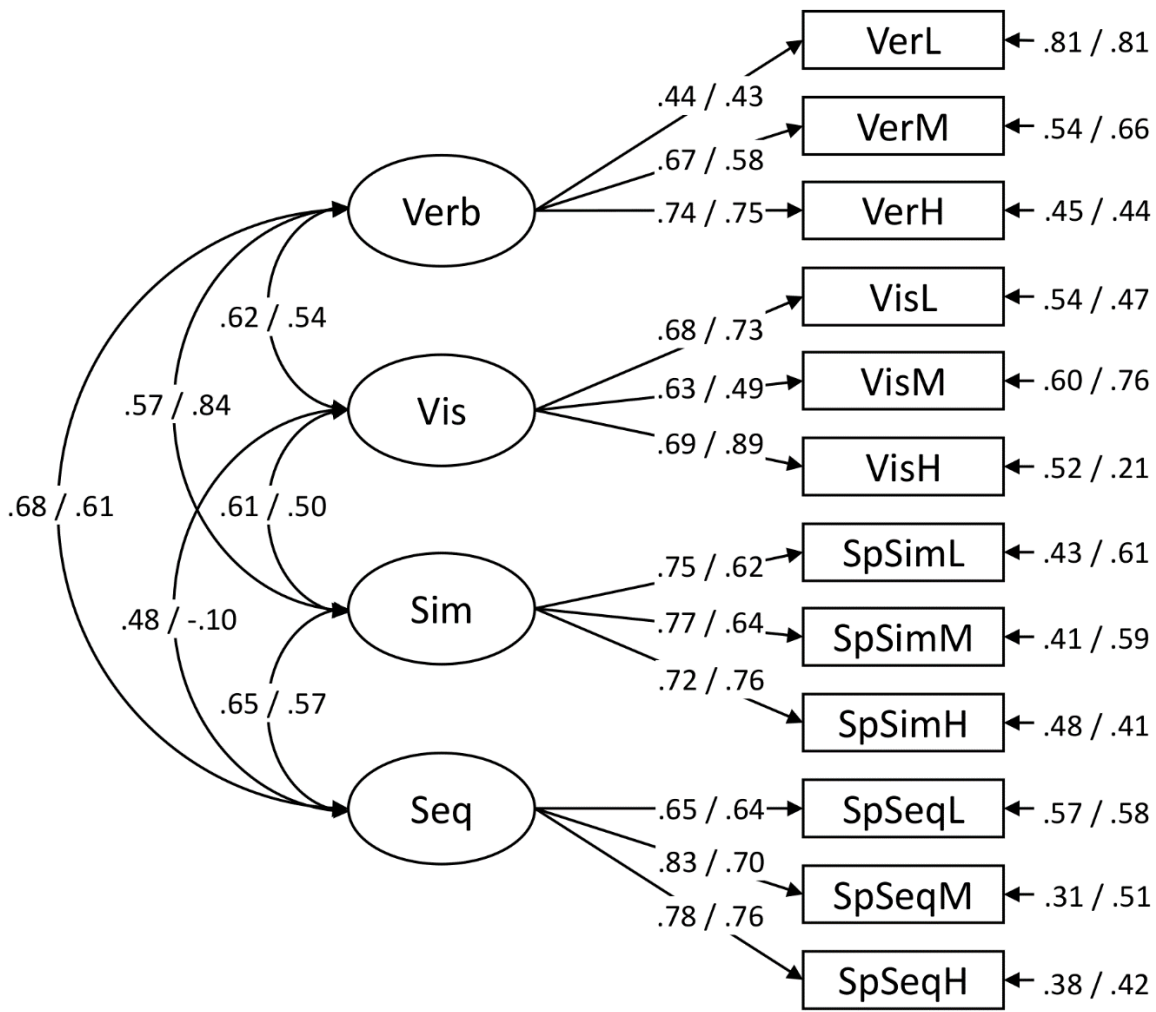

**Figure 2**. Standardized coefficients of the best-fitting multigroup CFA model (Model 4). For each parameter estimated, the coefficient on the left refers to the younger group, and the coefficient on the right to the older group. N = 739.

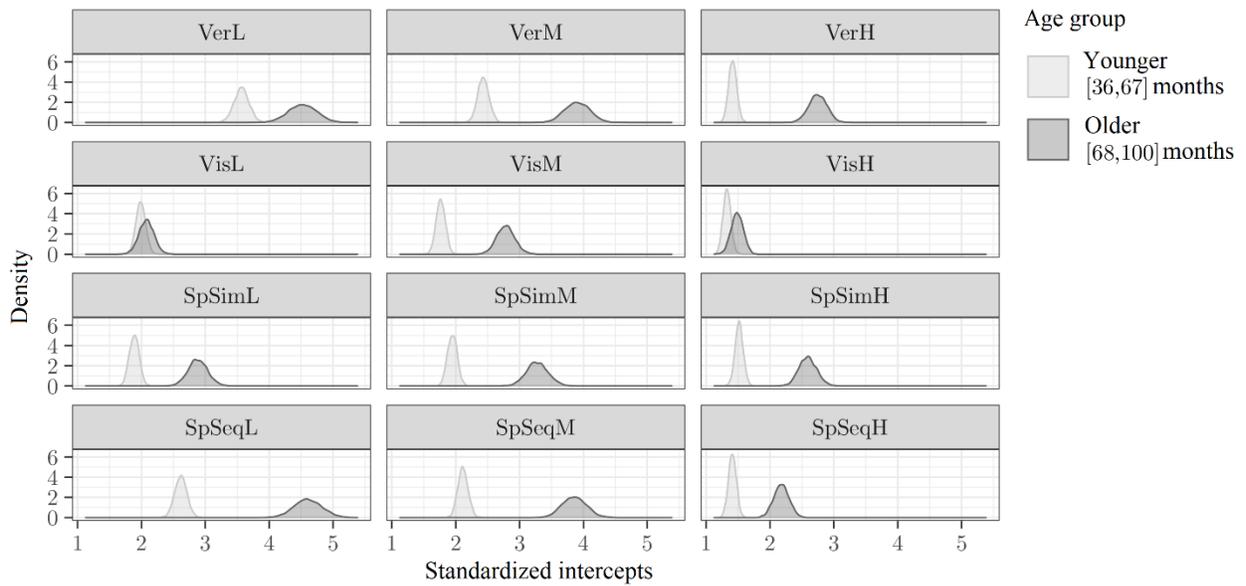

**Figure 3**. Posterior distributions of the standardized loadings of the best-fitting multigroup CFA model (Model 4), by age group.

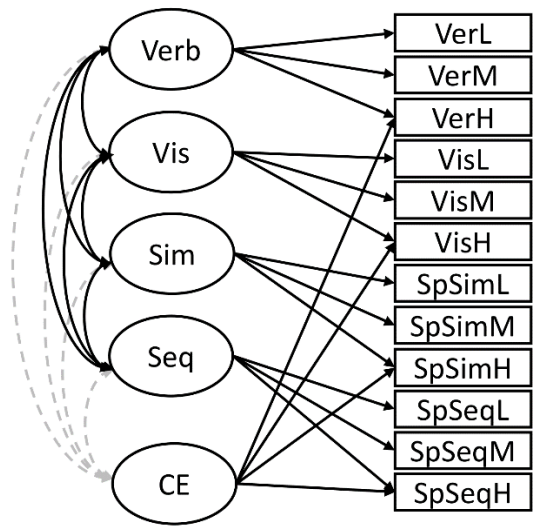

Figure 4. Graphical representation of Model 4-CE. The dashed gray correlations are set to zero.

**Table 1.** *Descriptions of the tasks used in the study. All the images used in the tasks were created by the authors.*

| Modality | Cognitive control | Material | Examples | Task |
|---|---|---|---|---|
| Verbal | Low | Word lists of increasing length | Cow, mum | Recall all the words |
| | Medium | Word lists of increasing length | House, flower | Recall the first words |
| | High | Word lists of increasing length | Mum, ball | Recall the first words and tap on the table when you hear the word "ball" |
| Spatial - sequential | Low | Matrixes of different sizes | 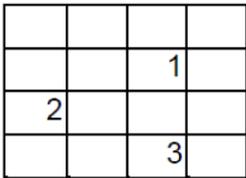 | Recall the sequence of steps taken by a frog |
| | Medium | Matrixes of different sizes | 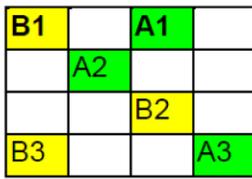 | Recall the first step of a sequence taken by one or two frogs |
| | High | Matrixes of different sizes | 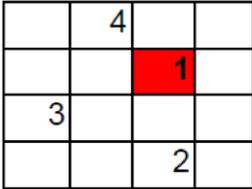 | Recall the first step of a sequence taken by a frog and tap your hand on the table when the frog jumps on a colored cell |
| Spatial - simultaneous | Low | Matrixes of different sizes | 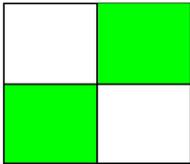 | Recall the position of the colored cells |
| | Medium | Matrixes of different sizes | 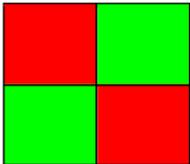 | Recall the position of either the green or the red cells |

| | High | Matrixes of different sizes | 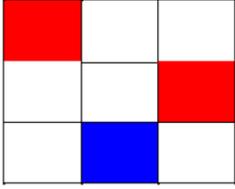 | Recall the position of the red cells and tap your hand on the table every time you see a blue cell |
|---|---|---|---|---|
| Visual | Low | Series of pictures of small houses | 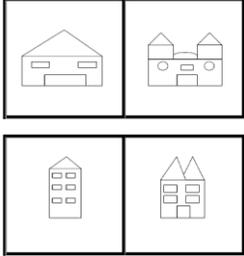 | Recall the order of presentation of the houses |
| | Medium | Series of pictures of small houses | 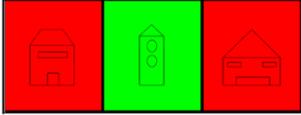 | Recall the position of the houses of a given color (either green or red) |
| | High | Series of pictures of small houses | 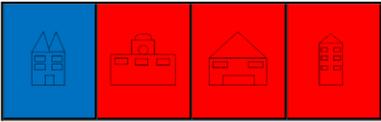 | Recall the position of the red houses and tap your hand on the table every time you see a blue house |

**Table 2** Fit indices of the multigroup CFA models, and of the same models fitted separately in the two age groups.

| Sample | Model | N | CC | BCFI | BTLI | BRMSEA | WAIC | Akaike weights |
|---|---|---|---|---|---|---|---|---|
| Multigroup | Model 1 | 739 | 100 | 0.73 | 0.65 | 0.14 | 34,136 | 0% |
| Multigroup | Model 2 | 739 | 100 | 0.75 | 0.67 | 0.14 | 34,075 | 0% |
| Multigroup | Model 3 | 739 | 100 | 0.84 | 0.78 | 0.11 | 33,820 | 0% |
| Multigroup | Model 4 | 739 | 100 | 0.95 | 0.93 | 0.07 | 33,516 | 100% |
| Multigroup | Model 5 | 739 | 100 | 0.92 | 0.83 | 0.10 | 33,631 | 0% |
| Multigroup | Model 6 | 739 | 100 | 0.93 | 0.88 | 0.08 | 33,584 | 0% |
| Multigroup | Model 7 | 739 | 100 | 0.94 | 0.89 | 0.08 | 33,570 | 0% |
| Younger | Model 1 | 501 | 100 | 0.76 | 0.69 | 0.14 | 23,241 | 0% |
| Younger | Model 2 | 501 | 100 | 0.79 | 0.72 | 0.13 | 23,182 | 0% |
| Younger | Model 3 | 501 | 100 | 0.86 | 0.80 | 0.11 | 23,052 | 0% |
| Younger | Model 4 | 501 | 100 | 0.95 | 0.92 | 0.07 | 22,866 | 100% |
| Younger | Model 5 | 501 | 100 | 0.90 | 0.77 | 0.12 | 22,976 | 0% |
| Younger | Model 6 | 501 | 100 | 0.94 | 0.88 | 0.08 | 22,896 | 0% |
| Younger | Model 7 | 501 | 100 | 0.94 | 0.90 | 0.08 | 22,885 | 0% |
| Older | Model 1 | 238 | 100 | 0.55 | 0.55 | 0.16 | 10,895 | 0% |
| Older | Model 2 | 238 | 100 | 0.55 | 0.55 | 0.16 | 10,894 | 0% |
| Older | Model 3 | 238 | 100 | 0.73 | 0.73 | 0.12 | 10,768 | 0% |
| Older | Model 4 | 238 | 100 | 0.94 | 0.94 | 0.06 | 10,648 | 92% |
| Older | Model 5 | 238 | 100 | 0.96 | 0.93 | 0.06 | 10,653 | 8% |
| Older | Model 6 | 238 | 100 | 0.92 | 0.86 | 0.09 | 10,689 | 0% |
| Older | Model 7 | 238 | 100 | 0.92 | 0.86 | 0.09 | 10,685 | 0% |

*Note*. Fit indices of the models considered. Median values based on 2,000 MCMC posterior samples are reported as estimates. CC = Convergence Criterion (a value less than 100 means the model does not converge); BCFI = Bayesian Comparative Fit Index; BTLI = Bayesian Tucker-Lewis Index; BRMSEA = Bayesian Root Mean Square Error of Approximation; WAIC = Widely Applicable Information Criterion.

Table 3 Standardized loadings and variances explained for Model 4CE in younger and older children.

| Group: YOUNGER | CE | | Ver | | Vis | | SpSim | | SpSeq | | | |
|---|---|---|---|---|---|---|---|---|---|---|---|---|
| Subtest | b | Var | b | Var | b | Var | b | Var | b | Var | h² | u² |
| VerL |  |  | .430 | .185 |  |  |  |  |  |  | .185 | .815 |
| VerM |  |  | .685 | .469 |  |  |  |  |  |  | .469 | .531 |
| VerH | .138 | .019 | .740 | .548 |  |  |  |  |  |  | .567 | .433 |
| VisL |  |  |  |  | .681 | .464 |  |  |  |  | .464 | .536 |
| VisM |  |  |  |  | .631 | .398 |  |  |  |  | .398 | .602 |
| VisH | .200 | .040 |  |  | .684 | .468 |  |  |  |  | .508 | .492 |
| SpSimL |  |  |  |  |  |  | .750 | .563 |  |  | .563 | .438 |
| SpSimM |  |  |  |  |  |  | .774 | .599 |  |  | .599 | .401 |
| SpSimH | .163 | .027 |  |  |  |  | .718 | .516 |  |  | .027 | .973 |
| SpSeqL |  |  |  |  |  |  |  |  | .650 | .423 | .423 | .578 |
| SpSeqM |  |  |  |  |  |  |  |  | .837 | .701 | .701 | .299 |
| SpSeqH | .131 | .017 |  |  |  |  |  |  | .783 | .613 | .630 | .370 |
| % total variance |  | 1.0 |  | 12.0 |  | 13.3 |  | 16.8 |  | 17.4 | 6.5 | 39.5 |
| % common variance |  | 1.7 |  | 19.9 |  | 22.0 |  | 27.7 |  | 28.7 |  |  |
| Group: OLDER | CE | | Ver | | Vis | | SpSim | | SpSeq | | | |
| Subtest | b | Var | b | Var | b | Var | b | Var | b | Var | h² | u² |
| VerL |  |  | .413 | .171 |  |  |  |  |  |  | .171 | .829 |
| VerM |  |  | .598 | .358 |  |  |  |  |  |  | .358 | .642 |
| VerH | .054 | .003 | .754 | .569 |  |  |  |  |  |  | .571 | .429 |
| VisL |  |  |  |  | .720 | .518 |  |  |  |  | .518 | .482 |
| VisM |  |  |  |  | .494 | .244 |  |  |  |  | .244 | .756 |
| VisH | -.144 | .021 |  |  | .908 | .824 |  |  |  |  | .845 | .155 |
| SpSimL |  |  |  |  |  |  | .610 | .372 |  |  | .372 | .628 |
| SpSimM |  |  |  |  |  |  | .644 | .415 |  |  | .415 | .585 |
| SpSimH | .046 | .002 |  |  |  |  | .778 | .605 |  |  | .002 | .998 |
| SpSeqL |  |  |  |  |  |  |  |  | .639 | .408 | .408 | .592 |
| SpSeqM |  |  |  |  |  |  |  |  | .713 | .508 | .508 | .492 |
| SpSeqH | .101 | .010 |  |  |  |  |  |  | .768 | .590 | .600 | .400 |
| % total variance |  | .4 |  | 11.0 |  | 15.9 |  | 13.9 |  | 15.1 | 56.2 | 43.8 |
| % common variance |  | .6 |  | 19.5 |  | 28.2 |  | 24.8 |  | 26.8 |  |  |

Note. h² = communality; u² = uniqueness.

**Supplemental Material**

**The structure of working memory in children from 3 to 8 years of age**

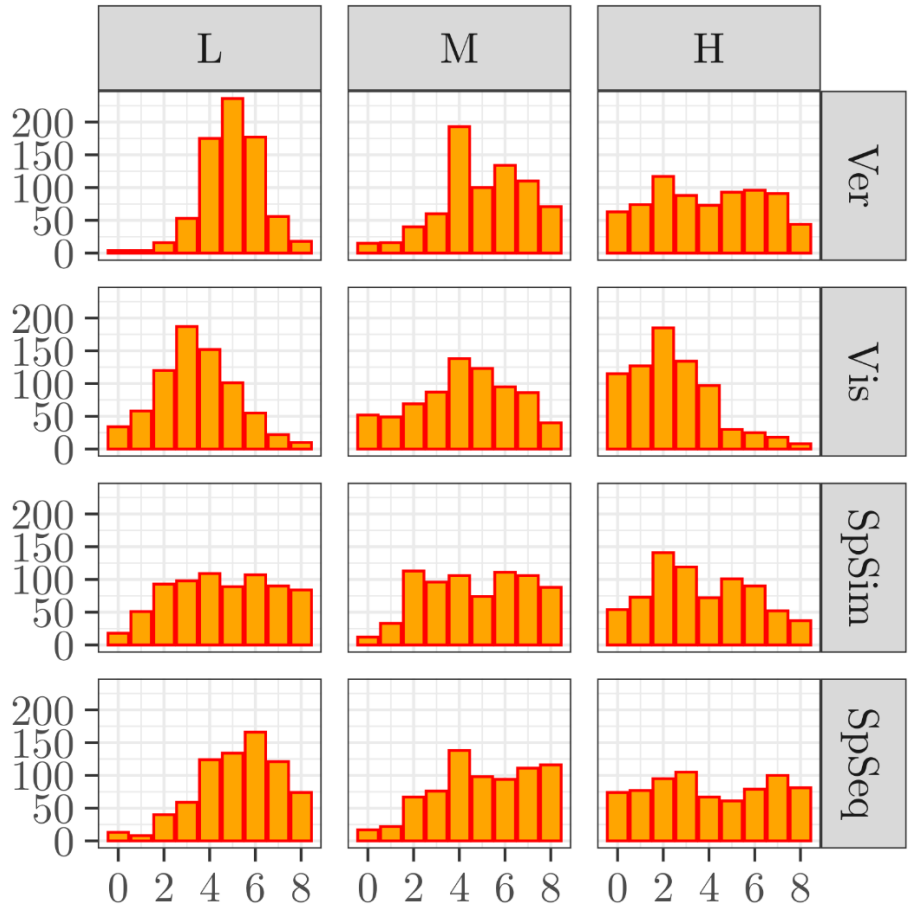

**Figure S1**. Univariate distributions in each task in the study (N = 739). Different modalities are presented on different rows of the plot (Ver = verbal, Vis = visual, SpSim = spatial simultaneous, SpSeq = spatial sequential); different levels of executive control are presented on different columns (L = low, M = medium, H = high).

**Table S1**

Descriptive statistics and correlations for the younger group (n = 501; 3.0-5.6 years of age) are presented below the main diagonal and at the bottom of the table. Descriptive statistics and correlations for the older group (n = 238; 5.7-8.3 years of age) are presented above the main diagonal and on the right of the table.

| Task | 1. | 2. | 3. | 4. | 5. | 6. | 7. | 8. | 9. | 10. | 11. | 12. | M | SD | Skewness | Kurtosis |
|---|---|---|---|---|---|---|---|---|---|---|---|---|---|---|---|---|
| 1. VerL | - | .282 | .337 | .193 | .192 | .316 | .209 | .319 | .249 | .048 | .043 | .140 | 5.25 | 1.15 | .03 | 3.13 |
| 2. VerM | .337 | - | .419 | .223 | .316 | .255 | .216 | .341 | .424 | .196 | .278 | .339 | 6.03 | 1.53 | -.52 | 2.49 |
| 3. VerH | .336 | .480 | - | .262 | .369 | .313 | .451 | .368 | .495 | .264 | .385 | .387 | 5.39 | 1.95 | -.76 | 2.88 |
| 4. VisL | .124 | .304 | .277 | - | .307 | .658 | .198 | .213 | .214 | -.144 | -.072 | -.094 | 3.85 | 1.84 | .44 | 2.68 |
| 5. VisM | .266 | .324 | .327 | .406 | - | .424 | .287 | .277 | .395 | .095 | .142 | .095 | 5.45 | 1.95 | -.97 | 3.62 |
| 6. VisH | .153 | .288 | .306 | .487 | .434 | - | .312 | .305 | .312 | -.093 | .010 | -.115 | 3.10 | 2.06 | .43 | 2.53 |
| 7. SpSimL | .162 | .321 | .280 | .428 | .306 | .404 | - | .425 | .468 | .228 | .277 | .197 | 5.80 | 2.00 | -.73 | 2.52 |
| 8. SpSimM | .158 | .310 | .291 | .305 | .245 | .258 | .578 | - | .486 | .263 | .260 | .230 | 6.05 | 1.83 | -.91 | 2.88 |
| 9. SpSimH | .238 | .323 | .335 | .281 | .269 | .297 | .547 | .557 | - | .340 | .332 | .351 | 5.22 | 2.00 | -.64 | 2.64 |
| 10. SpSeqL | .221 | .266 | .307 | .213 | .245 | .175 | .269 | .438 | .345 | - | .475 | .474 | 6.29 | 1.35 | -1.04 | 4.90 |
| 11. SpSeqM | .158 | .374 | .416 | .264 | .323 | .202 | .355 | .465 | .400 | .562 | - | .536 | 6.38 | 1.64 | -1.02 | 3.32 |
| 12. SpSeqH | .171 | .374 | .517 | .237 | .344 | .253 | .335 | .417 | .345 | .485 | .654 | - | 5.40 | 2.46 | -.81 | 2.39 |
| M | 4.84 | 4.50 | 3.21 | 3.23 | 3.61 | 2.07 | 3.94 | 4.02 | 2.92 | 4.73 | 4.35 | 3.38 | | | | |
| SD | 1.35 | 1.84 | 2.26 | 1.62 | 2.04 | 1.56 | 2.08 | 2.06 | 1.92 | 1.80 | 2.05 | 2.38 | | | | |
| Skewness | -.39 | -.21 | .39 | -.07 | .00 | .75 | .14 | .24 | .52 | -.32 | .02 | .37 | | | | |
| Kurtosis | 3.88 | 2.84 | 2.14 | 2.51 | 2.39 | 3.78 | 2.19 | 2.17 | 2.75 | 2.88 | 2.4 | 2.11 | | | | |

**Age trends**

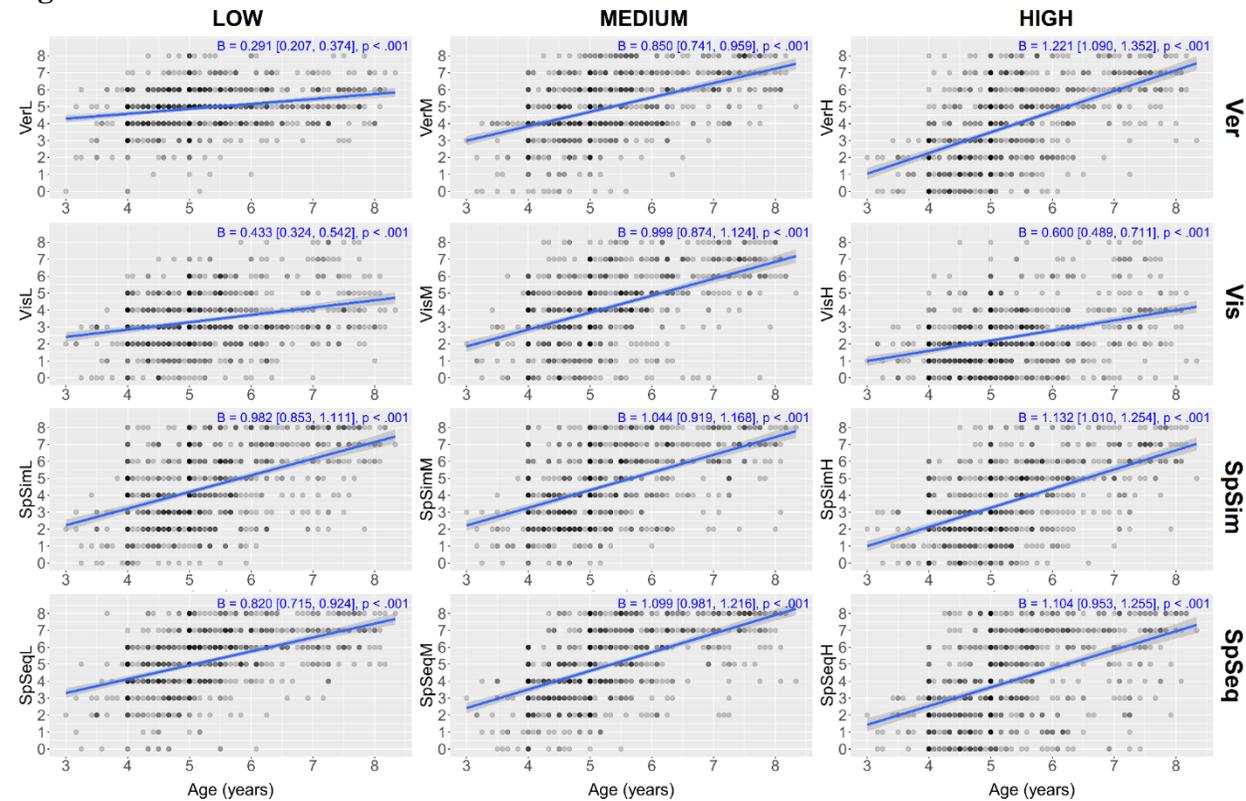

**Figure S2.** Scatter plots and linear models representing the age trends of the performance in all 12 tasks. Shaded areas represent 95% confidence bands. Values at Age = 3 years can be represented as the intercepts for our sample and compared across tasks. The B coefficients reported in blue are the raw slope coefficients and can be compared across tasks.

# Standardized intercepts

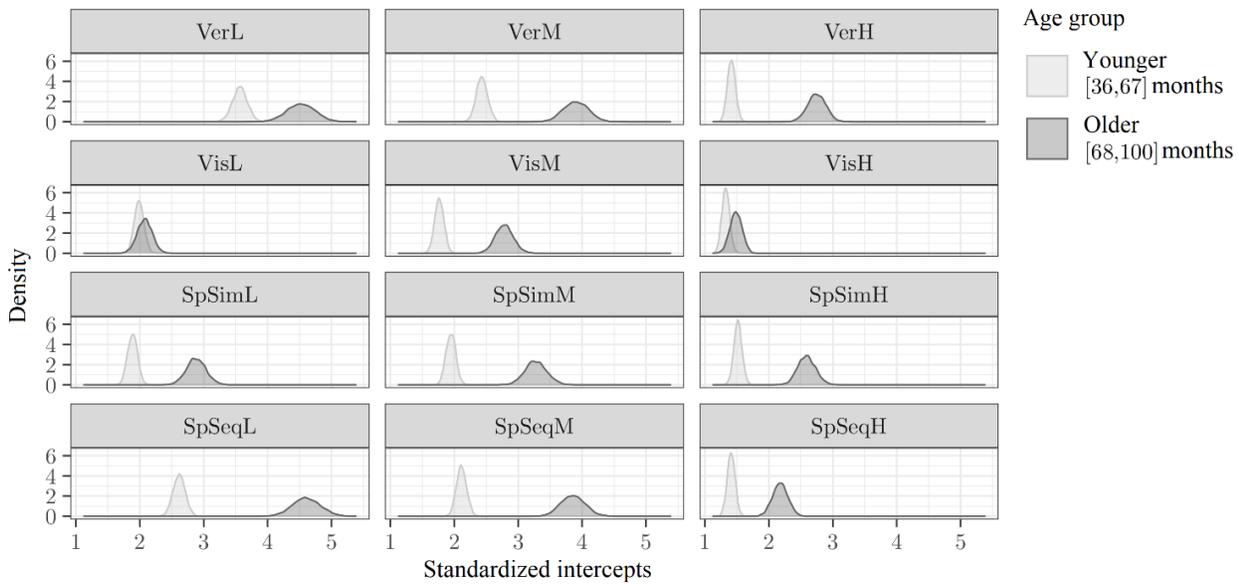

**Figure S3**. Posterior distributions of the tasks standardized intercepts of tasks, separately by age group, in the best fitting multigroup CFA model (Model 4).

**Table S2**.

Estimated standardized intercepts in the two groups, estimated overlapping between their posterior distributions, and average differences (Δ) with 90% HPDI.

| Task | Group | | Overlapping | Δ | 90% HPDI |
|---|---|---|---|---|---|
| | Younger | Older | | | |
| VerL | 3.57 | 4.52 | 0% | .96 | (.53, 1.35) |
| VerM | 2.43 | 3.90 | 0% | 1.47 | (1.12, 1.82) |
| VerH | 1.42 | 2.75 | 0% | 1.33 | (1.08, 1.58) |
| VisL | 1.99 | 2.08 | 45% | .09 | (-.013, .32) |
| VisM | 1.77 | 2.78 | 0% | 1.02 | (.74, 1.27) |
| VisH | 1.33 | 1.49 | 19% | .16 | (-.03, .34) |
| SpSimL | 1.89 | 2.88 | 0% | 1.00 | (.72, 1.26) |
| SpSimM | 1.95 | 3.28 | 0% | 1.33 | (1.04, 1.65) |
| SpSimH | 1.52 | 2.59 | 0% | 1.08 | (.84, 1.33) |
| SpSeqL | 2.62 | 4.61 | 0% | 1.99 | (1.62, 2.39) |
| SpSeqM | 2.11 | 3.86 | 0% | 1.74 | (1.40, 2.08) |
| SpSeqH | 1.41 | 2.18 | 0% | .77 | (.55, .99) |

*Note*. HPDI = highest posterior density interval.

**Standardized loadings**

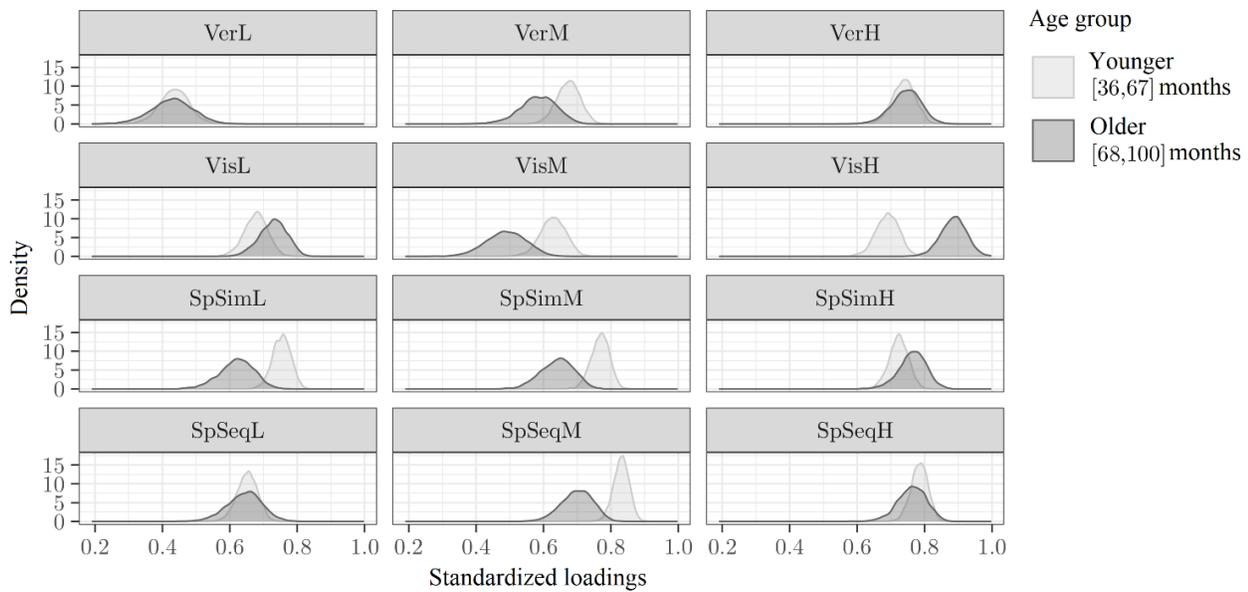

**Figure S4**. Posterior distributions of the standardized loadings of tasks, separately by age group, in the best fitting multigroup CFA model (Model 4). This figure is identical to Figure 3 in the manuscript, and it is reported here only for completeness.

**Table S3**.

Estimated standardized loadings in the two groups, estimated overlapping between their posterior distributions, and average differences (Δ) with 90% HPDI.

| Task | Group | | Overlapping | Δ | 90% HPDI |
|---|---|---|---|---|---|
| | Younger | Older | | | |
| VerL | .44 | .43 | 70% | -.01 | (-.13, .12) |
| VerM | .67 | .58 | 20% | -.09 | (-.20, .01) |
| VerH | .74 | .75 | 74% | .01 | (-.08, .11) |
| VisL | .68 | .73 | 33% | .05 | (-.04, .14) |
| VisM | .63 | .49 | 9% | -.14 | (-.25, -.02) |
| VisH | .69 | .89 | 1% | .20 | (.12, .28) |
| SpSimL | .75 | .62 | 5% | -.13 | (-.23, -.04) |
| SpSimM | .77 | .64 | 6% | -.13 | (-.22, -.04) |
| SpSimH | .72 | .76 | 35% | .04 | (-.04, .12) |
| SpSeqL | .65 | .64 | 61% | -.01 | (-.11, .08) |
| SpSeqM | .83 | .70 | 4% | -.13 | (-.21, -.04) |
| SpSeqH | .78 | .76 | 52% | -.02 | (-.10, .06) |

*Note*. HPDI = highest posterior density interval.

**Factor correlations (standardized covariances)**

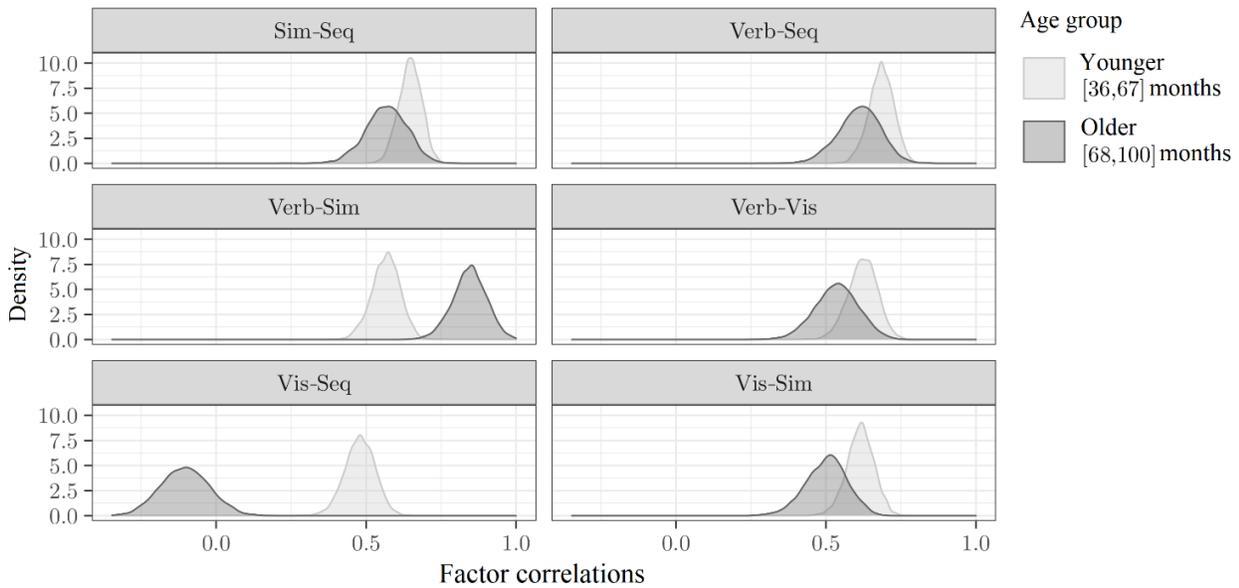

**Figure S5**. Posterior distributions of the correlations between factors, separately by age group, in the best fitting multigroup CFA model (Model 4).

**Table S4**.

Estimated factor correlations in the two groups, estimated overlapping between their posterior distributions, and average differences (Δ) with 90% HPDI.

| Correlation | Group | | Overlapping | Δ | 90% HPDI |
|---|---|---|---|---|---|
| | Younger | Older | | | |
| Sim-Seq | .65 | .57 | 31% | -.09 | (-.22, .06) |
| Verb-Seq | .68 | .61 | 34% | -.07 | (-.21, .06) |
| Verb-Sim | .57 | .84 | 1% | .28 | (.15, .39) |
| Verb-Vis | .62 | .54 | 19% | -.11 | (-.25, .02) |
| Vis-Seq | .48 | -.10 | 0% | -.58 | (-.74, -.43) |
| Vis-Sim | .61 | .50 | 30% | -.08 | (-.21, .05) |

*Note*. HPDI = highest posterior density interval.

**Standardized residual variances**

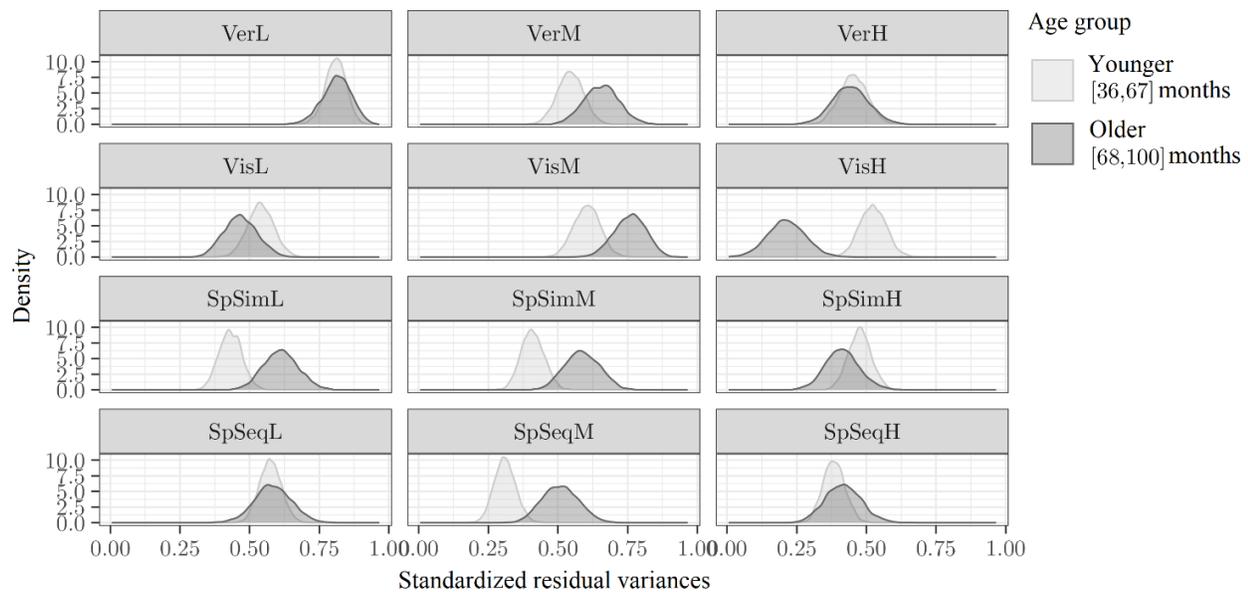

**Figure S6**. Posterior distributions of the standardized residual variances of the tasks, separately by age group, in the best fitting multigroup CFA model (Model 4).

**Table S5**.

Estimated standardized residual variances in the two groups, estimated overlapping between their posterior distributions, and average differences (Δ) with 90% HPDI.

| Task | Group | | Overlapping | Δ | 90% HPDI |
|---|---|---|---|---|---|
| | Younger | Older | | | |
| VerL | .81 | .81 | 70% | .00 | (-.10, .11) |
| VerM | .54 | .66 | 20% | .11 | (-.02, .24) |
| VerH | .45 | .44 | 74% | -.01 | (-.15, .11) |
| VisL | .54 | .47 | 33% | -.07 | (-.20, .05) |
| VisM | .60 | .76 | 9% | .15 | (.04, .29) |
| VisH | .52 | .21 | 1% | -.31 | (-.44, -.18) |
| SpSimL | .43 | .61 | 5% | .18 | (.06, .30) |
| SpSimM | .41 | .59 | 6% | .18 | (.05, .30) |
| SpSimH | .48 | .41 | 35% | -.06 | (-.18, .06) |
| SpSeqL | .57 | .58 | 61% | .01 | (-.11, .13) |
| SpSeqM | .31 | .51 | 4% | .20 | (.07, .32) |
| SpSeqH | .38 | .42 | 52% | .03 | (-.09, .15) |

*Note*. HPDI = highest posterior density interval.

**Additional analysis**

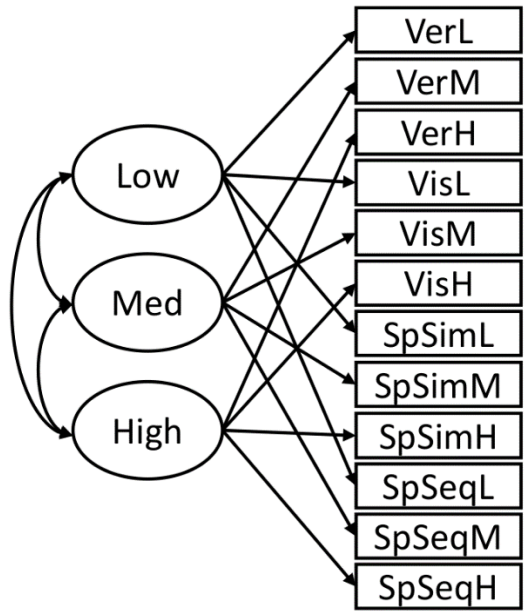

**Figure S7**. Additional model featuring three correlated factors representing low, medium, and high control levels, but without modality-specific factors.

The model shown in Figure S7 has inadequate fit at the multi-group level, as well as in the younger and older groups separately. For the multi-group model: BCFI = 0.73, BTLI = 0.64, BRMSEA = 0.14, WAIC = 34,147. For the younger group: BCFI = 0.76, BTLI = 0.68, BRMSEA = 0.14, WAIC = 23,247. For the older group: BCFI = 0.65, BTLI = 0.55, BRMSEA = 0.16, WAIC = 10,900.